\documentclass[aps,preprint]{revtex4}
\def\beq{\begin{equation}} \def\eeq{\end{equation}}
\usepackage{graphics,epsfig,amssymb,multirow}

\begin{document}

\title{Isovector and isoscalar pairing in odd-odd $N=Z$ nuclei within a quartet approach}

\author{ D. Negrea and N. Sandulescu }
\affiliation{ National Institute of Physics and Nuclear Engineering, 
	 76900 Bucharest-Magurele, Romania }

\author{D. Gambacurta}
\affiliation{ Extreme Light Infrastructure - Nuclear Physics (ELI-NP),
 76900 Bucharest-Magurele, Romania}
 
\begin{abstract}

The quartet condensation model (QCM) is extended for the treatment of isovector and isoscalar pairing
in odd-odd N=Z nuclei.  In the extended QCM approach the lowest states of isospin T=1 and T=0 in 
odd-odd nuclei are described variationally by trial functions composed by a proton-neutron pair 
appended to a condensate of  4-body operators. The latter are taken as a linear superposition
of an isovector quartet, built by two isovector pairs coupled to the total isospin T=0, and two collective isoscalar 
pairs.  In all pairs the nucleons are distributed in time-reversed single-particle states of axial symmetry.
The accuracy of the trial functions is tested for realistic pairing Hamiltonians and odd-odd N=Z nuclei 
with the valence nucleons moving above the cores $^{16}$O, $^{40}$Ca and $^{100}$Sn. It is shown that
the extended QCM approach is able to predict with high accuracy the energies of the lowest T=0 and T=1 states.
The present calculations indicate that in these states the isovector and the isoscalar pairing correlations
coexist together, with the former playing a dominant role.
\end{abstract}

\maketitle

\section{introduction}

Many experimental and theoretical studies have been dedicated lately to the role played by  
the isoscalar and isovector proton-neutron (pn) pairing in odd-odd N=Z nuclei (e.g., see 
\cite{fm,sagawa_review} and the references quoted therein). 
The experimental data show that the ground states of odd-odd  N=Z nuclei have the isospin 
T=0 for  $A < 34$  and, with some exceptions, the isospin T=1 for heavier nuclei. 
This fact is sometimes considered as an indication of the dominant role of  isoscalar (T=0) pn pairing 
in light N=Z nuclei. The fingerprints of T=0 pn pairing in odd-odd N=Z nuclei is also investigated 
lately  in relation to the Gamow-Teller (GT) charge-exchange reactions. Thus in some odd-odd N=Z nuclei
 there is  an enhancement of the GT strength in the low-energy region which appears to be 
 sensitive to the T=0 pn interaction \cite{fujita}. The competition between the isovector and isoscalar
 pairing in odd-odd nuclei was also discussed extensively in relation to the odd-even mass difference
 along N=Z line \cite{vogel,macchiavelli}.
  
 On theoretical side, the role of pn pairing in odd-odd N=Z nuclei is still not clear. 
A  fair description of low-lying states and GT  transitions in odd-odd N=Z nuclei is given by the
shell model (SM) calculations (e.g., see \cite{sm_gt}).  However, due to the complicated structure of the
SM wave function, from these calculations it is not easy to draw conclusions on the role played by the pn pairing. 
 Recently, the effect of T=0 and T=1 pairing forces on the spectroscopic properties
 of odd-odd N=Z nuclei was analyzed  in the framework of a simple three-body model in which the odd pn
  pair is supposed to move on the top of a closed even-even core  \cite{sagawa_odd}. 
  This model gives good results for the nuclei in which the core can be considered as inert, 
  such as $^{18}$F and $^{42}$Sc,  but not for the nuclei in which the core degrees of freedom are important.

The difficulties mentioned  above point to the need of new microscopic models which, on one hand, 
to be able to describe reasonably well the spectroscopic properties of  odd-odd N=Z nuclei,  and, 
on the other hand,  to be simple enough for understanding the impact of pn pairing correlations 
on  physical observables. As an alternative, in this article we shall use the framework of the quartet
condensation model (QCM) we have proposed in Ref. \cite{qcm_def}. Its advantage is the explicit treatment
of the pairing correlations in the wave function and, compared to other pairing models, 
the exact conservation of particle number and  isospin. The scope of this study is to
extend the QCM approach of Ref. \cite{qcm_def}, applied previously to even-even nuclei, 
for the case of odd-odd N=Z nuclei and to study, for these nuclei, the role played by
proton-neutron pairing in the lowest T=0 and T=1 states. 

\section{Formalism}
 
In the present study the isovector and isoscalar pairing correlations in odd-odd N=Z nuclei are
described by  pairing forces which act  on  pairs of nucleons moving in time-reversed 
single-particle states generated by axially-deformed mean fields. The corresponding 
Hamiltonian is given by
\beq
\hat{H}= \sum_{i,\tau=\pm 1/2} \varepsilon_{i\tau} N_{i\tau} +
 \sum_{i,j} V^{T=1}(i,j) \sum_{t=-1,0,1} P^+_{i,t} P_{j,t} +
 \sum_{i,j} V^{T=0}(i,j)  D^+_{i,0} D_{j,0}.\eeq
In the first term $\varepsilon_{i\tau}$ are the single-particle energies for the neutrons ($\tau=1/2$) and 
protons ($\tau=-1/2$) while $N_{i\tau}$ are the particle number operators. The second term is the isovector 
pairing interaction expressed by the pair operators  
$P^+_{i,0}=(\nu^+_i \pi^+_{\bar{i}} + \pi^+_i \nu^+_{\bar{i}})/\sqrt{2}$, 
$P^+_{i,1}=\nu^+_i \nu^+_{\bar{i}}$ and 
$P^+_{i,-1}=\pi^+_i \pi^+_{\bar{i}}$, where  $\nu^+_i$ and $\pi^+_i$ are creation operators for neutrons and  protons 
in the state $i$.  The last term is the  isoscalar pairing interaction represented by the operators 
$D^+_{i,0}=(\nu^+_i \pi^+_{\bar{i}} - \pi^+_i \nu^+_{\bar{i}})/\sqrt{2}$ which creates a non-collective 
isoscalar pair in the time reversed states $(i,\bar{i})$.  In the applications considered in 
the present paper the single-particle states have axial symmetry.

The Hamiltonian (1) was employed recently to study  the isovector and isoscalar pairing correlations in 
even-even N=Z nuclei in the framework of QCM approach \cite{qcm_def}. This approach is extended
here for the case of odd-odd nuclei. For consistency reason we start by presenting shortly 
the QCM approach for even-even nuclei.
 
In the QCM approach the ground state of the Hamiltonian (1) for a system of N neutrons and Z protons, with
N=Z=even, moving above a closed core $|0 \rangle$ is described by the ansatz
\begin{equation}
| QCM \rangle =(A^+ + (\Delta^+_0)^2)^{n_q} |0 \rangle,
\end{equation}
where $n_q=(N+Z)/4$.
The operator $A^+$ is the collective  quartet defined by a superposition of two non-collective
isovector pairs coupled to total isospin T=0 and has the expression
 \beq
A^+ = \sum_{i,j} x_{ij} [P^+_i P^+_j]^{T=0}.
\eeq
Supposing that the mixing amplitudes $x_{ij}$ are separable, that is $x_{ij}=x_i x_j$, the collective quartet
gets the form
\beq
A^+= 2 \Gamma^+_1 \Gamma^+_{-1} - (\Gamma^+_0)^2,
\eeq
where $\Gamma^+_{t}= \sum_i x_i P^+_{i,t}$ are the  collective neutron-neutron (t=1), proton-proton (t=-1)
and proton-neutron (t=0) isovector pairs. Finally, in Eq. (2) the operator $\Delta^+_0$ is the collective 
isoscalar pair defined by
\beq
\Delta^+_{0}= \sum_i y_i D^+_{i,0}.
\eeq

When the single-particle states are degenerate and the strength of the two pairing forces are equal,
the  QCM state (2) is the  exact solution of the Hamiltonian (1). For realistic single-particle spectra
and realistic pairing interactions the QCM state (2) is not anymore the exact solution but, as shown
in Ref. \cite{qcm_def}, it predicts with high accuracy the pairing correlations in even-even N=Z nuclei.

In what follows we extend the QCM approach to odd-odd N=Z systems. 
The main assumption, suggested by the exact solution of the Hamiltonian (1) (see below),
is that the lowest T=1 and T=0 states in odd-odd nuclei can be well described variationally
by trial states  obtained by appending to the  QCM function (2) a proton-neutron pair. 
Since the isospin of the QCM state (2) is T=0, the total isospin of the odd-odd system is 
given by the isospin of the appended pair.
Thus, the ansatz for the lowest T=1 state of the odd-odd N=Z systems is
\begin{equation}
| iv;QCM \rangle = \tilde{\Gamma}^+_0 (A^+ + (\Delta^+_0)^2)^{n_q} |0 \rangle,
\end{equation}
where $\tilde{\Gamma}^+_0=\sum_i z_i P^+_{i,0}$ is the isovector pn pair attached to the
the even-even part of the state (in what follows we shall use the name "core" for the even-even
part of the state (6), which should be not confused with the closed core $|0 \rangle$). 
It can be seen that this pair has a different collectivity compared to the isovector pn 
pair $\Gamma^+_0$ contained  in the quartet $A^+$ (see Eq. 4).

Likewise, the lowest T=0 state of odd-odd N=Z systems is described by the  function
\begin{equation}
| is;QCM \rangle = \tilde{\Delta}^+_0 (A^+ + (\Delta^+_0)^2)^{n_q} |0 \rangle,
\end{equation}
where $\tilde{\Delta}^+_0=\sum_i z_i D^+_{i,0}$ is the odd  isoscalar pair, which
has also a different structure compared to the isoscalar  pair $\Delta^+_0$ which
enters in the even-even core. Due to its different isospin, the state (7) is orthogonal
to the isovector state (6).

We have proved that the states (6,7) are the exact eigenfunctions of the Hamiltonian (1) when 
the single-particle energies are degenerate and when the pairing forces have the same strength, i.e.,
$V^{T=1}(i,j) = V^{T=0}(i,j)=g$. In this case the  states (6,7) have the same energy which, 
for  $\epsilon_i=0$, is given by
\beq
E(n_q,\nu)=(\nu-2n_q)g+2n_q(\nu-n_q+2)g,
\eeq
where $n_q$ is the number of quartets and $\nu$ is the number of single-particle 
levels. In Eq. (8) the second term corresponds to the energy of the even-even core of the
functions (6,7). It is worth to be mentioned that this exact solution is not the exact solution of the
SU(4) model \cite{Dobes} because in the Hamiltonian (1) the isoscalar force contains only
pairs in time-reversed single-particle states.

For a non-degenerate single-particle spectrum and general pairing forces the QCM states (6,7)
are determined variationally. The variational parameters  are the amplitudes $x_i$, $y_i$ and $z_i$ which  are 
defining, respectively, the  isovector pairs $\Gamma^+_t$, the isoscalar pair $\Delta^+_0$ and the
odd pn pair. They are found by minimizing the average of Hamiltonian (1) on the QCM states (6,7) 
and by imposing, for the latter, the normalization condition. The average of the Hamiltonian and the
norm of the QCM states are calculated using the technique of reccurence relations. More precisely,
the calculations are performed using auxiliary states composed by products of collective pairs.
Thus, for the isovector T=1 state (6) the auxiliary states are                        
\beq
| n_1 n_2 n_3 n_4 n_5 \rangle = (\Gamma_1^+)^{n_1} (\Gamma_{-1}^+)^{n_2} (\Gamma_0^+)^{n_3} (\Delta_0^+)^{n_4} (\tilde{\Gamma}_0^+)^{n_5} |0\rangle.
\eeq
The auxiliary states for the calculations of isoscalar T=0 state (7) have a similar structure with the 
difference that the odd isovector pair $\tilde{\Gamma}_0^+$ is replaced by the odd isoscalar
pair $\tilde{\Delta}_0^+$. It can be observed  that the QCM states (6,7) can be  expressed in
terms of a subset of auxiliary states corresponding to  specific combinations of $n_i$.  
However, in order to close the 
recurrence relations one needs to evaluate the matrix elements of the Hamiltonian (1) for  all 
auxiliary states which satisfy the conditions $\sum_i n_i=(N+Z)/2$ and $n_5=0,1$. An example of recurrence relations, for the case
of even-even systems, can be seen in  Refs. \cite{qcm_iv,thesis}. 

The advantage of the QCM approach is the possibility to investigate in a direct manner
the role of various types of correlations by simply switching them on and off in the structure
of the states (6,7).  
Thus, in order to explore the  importance of isoscalar pairing on the lowest  T=0 and T=1 states in odd-odd N=Z systems
one can remove from the functions (6,7)  the  isoscalar pair $\Delta^+_0$.
In this approximation the functions get the  expressions
\begin{equation}
| is;Q_{iv} \rangle = \tilde{\Delta}^+_0 (A^+)^{n_q} |0 \rangle,
\end{equation}
\begin{equation}
| iv;Q_{iv} \rangle = \tilde{\Gamma}^+_0 (A^+)^{n_q} |0 \rangle.
\end{equation}
Alternatively, we can estimate the importance of the isovector pairing by removing  from 
the QCM functions the  isovector quartet $A^+$. The corresponding functions are
\begin{equation}
| C_{is} \rangle = (\Delta^+_0)^{2n_q+1} |0 \rangle,
\end{equation}
\begin{equation}
| iv; C_{is} \rangle = \tilde{\Gamma}^+_0 (\Delta^{+2}_0)^{n_q} |0 \rangle.
\end{equation} 
Another possibility is to remove from the QCM functions  the contribution of like-particle pairs, keeping
only the isovector and isoscalar pn pairs. These trial states, which can be employed to study
the role of like-particle pairing in N=Z nuclei, have the expressions
\begin{equation}
| is;C_{iv} \rangle = \tilde{\Delta}^+_0 (\Gamma^{+2}_0)^{n_q} |0 \rangle,
\end{equation}
\begin{equation}
| C_{iv} \rangle = (\Gamma^+_0)^{2n_q+1} |0 \rangle.
\end{equation} 
Contrary to the previous approximations, 
the states (14,15) have not a well-defined isospin. 

Among the approximations mentioned above of special interest are the ones corresponding
to the states (12) and (15), which are pure  condensates of isoscalar and, respectively, 
isovector pn pairs. These states are sometimes considered as representative for understanding
the competition between   isovector and isoscalar proton-neutron paring in nuclei.

The QCM states (6,7) and all the approximations based on them are formulated here in the intrinsic
system associated to the axially deformed single-particle levels. 
Therefore they have a well-defined projection
of the angular momentum on z-axis but not a well-defined angular momentum. A more complicated
quartet formalism for odd-odd nuclei, which conserves exactly the  angular momentum and 
takes into account the correlations induced by a general two-body force, was proposed recently
in Ref. \cite{sasa_odd}.

\section{Results}  

To test the accuracy of  the QCM approach for odd-odd N=Z nuclei  we consider nuclei having  
protons and neutrons outside the closed cores $^{16}$O, $^{40}$Ca and $^{100}$Sn. The calculations
are performed employing for  the pairing forces and the single-particle states a similar input  as in our
previous study for even-even nuclei \cite{qcm_def}. Thus, the single-particle states are generated 
by axially deformed mean fields calculated with the Skyrme-HF code $ev8$ \cite{ev8} and  with 
the force $Sly4$ \cite{sly4}. In the mean field calculations the Coulomb interaction is switched off,
so the single-particle energies for protons and neutrons are the same.
For the pairing forces we use  a zero range delta interaction $V^T(r_1,r_2)=V_0^T\delta(r_1-r_2)
\hat{P}^T_{S,S_z}$,
where $\hat{P}^T_{S,S_z}$ is the projection operator on the spin of the pairs, i.e., $S=0$ for the isovector (T=1)
force and  $S=1,S_z=0$ for the isoscalar (T=0) force. The matrix elements of the pairing forces are calculated 
using the single-particle wave functions generated by the Skyrme-HF calculations (for details, see \cite{gambacurta}).
As parameters we use the strength of the isovector
force, denoted by $V_0$, and the  scaling factor $w$ which defines the strength of the isoscalar force,
$V_0^{T=0}=w V_0$. How to fix these parameters is not a simple task. Since the main goal of this study is
to test the accuracy of the QCM approach,  we have made several calculations with various strengths,  
 $V_0=\{300, 465, 720, 1000\}$ MeV fm$^{-3}$, which cover all possible situations, from the weak to
 the strong pairing regime. Because the conclusions relevant for this study are quite similar for all these strengths,
here we are presenting only the results  for the pairing strength $V_0=465$ MeV fm$^{-3}$ employed  in our previous 
study of even-even nuclei \cite{qcm_def}.

\begin{figure}[h]
\centering
\includegraphics[trim = 0.04cm 0.50cm 0.05cm 0.4cm,clip=true, width=10cm]
{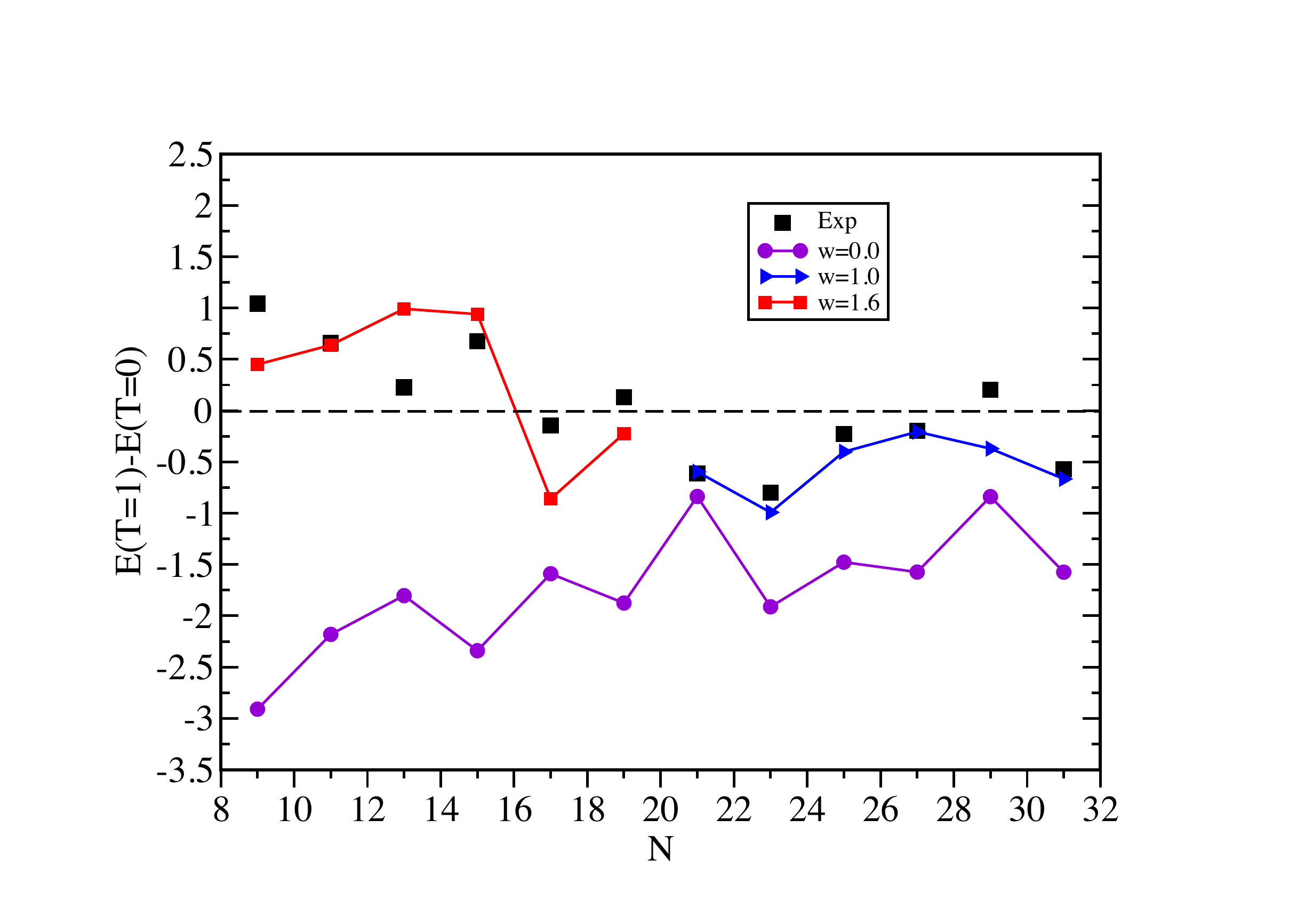}
\caption{The energy difference between the lowest T=1 and T=0 states as a function of N=Z=A/2. The experimental data are
 extracted from Ref. \cite{data}. The solid lines show the exact results obtained by diagonalising the Hamiltonian (1). The calculations
 correspond to the  strength $V_0$=465 MeV fm$^{-3}$ and to various scaling factors $w$.} 
\end{figure}

For the scaling factor $w$ we also used various  values, $w=\{1.0,1.3,1.5,1.6\}$. 
To find the most appropriate value of $w$ for the strength $V_0$= 465 MeV fm$^{-3}$ we
have searched for the best agreement with the energy difference between the first excited 
state and the ground state of odd-odd nuclei. These energy differences are shown in Fig. 1
by black squares.
 It is worth mentioning that  the lowest T=0 state can have various 
angular momenta $J \ge1$ (e.g., the ground states of $^{22}$Na and $^{26}$Al have $J=3$ and,
respectively, $J=5$). The theoretical results shown in Fig. 1 corresponds to the exact
diagonalization of  Hamiltonian (1) in a space spanned by 10 single-particle levels above 
the cores $^{16}$O and $^{40}$Ca.
The best agreement with the experimental data is obtained by choosing $w=1.6$ for $sd$-shell nuclei and 
$w=1.0$ for $pf$-shell nuclei.  As seen in Fig. 1, for these parameters the calculations predict rather 
well how the isospin of the ground state is changing with the mass number.   Since for the nuclei above $^{100}$Sn
there are no available experimental data on low-lying states which to be used for fixing
the scaling factor $w$, in the calculations presented below we have chosen for $w$ the same 
value as for the pf-shell nuclei. In Fig. 1 we show also the results obtained considering only the 
isovector pairing force, that is, for
$w=0.0$. It can be seen that in this case the predictions are quite far from the data, especially
for the $sd$-shell nuclei.

\begin{table}[ht]
	\caption{Correlation energies, in MeV, for the lowest T=1 and T=0 states. In the brackets are given the errors relative
	to the exact values indicated in the 3rd column. Are shown the results corresponding to the  QCM states (6,7) and 
	to the approximations defined by Eqs. (10-15).}	
	\begin{center}
		\begin{tabular}{|c|c|c|c|c|c|c|}
			
			\hline
			\hline
			
			& 
			& Exact 
			& $\vert QCM \rangle$ 
			& $\vert iv;QCM_{iv} \rangle/ \vert is;QCM_{iv} \rangle$ 
			& $\vert iv;C_{is} \rangle/ \vert C_{is} \rangle$  
			& $\vert C_{iv} \rangle/ \vert is;C_{iv} \rangle$  \\
			
			\hline
			\hline
			
			$^{ 22}$Na & T=0 & 13.87 & 13.87 (0.00\%) & 13.86 (0.07\%) & 13.85 (0.12\%) & 13.85 (0.15\%) \\
			\hline
			$$         & T=1 & 13.23 & 13.23 (0.03\%) & 13.22 (0.05\%) & 12.97 (1.97\%) & 13.22 (0.11\%) \\
			\hline
			$^{ 26}$Al & T=0 & 22.06 & 22.05 (0.03\%) & 22.04 (0.07\%) & 21.94 (0.53\%) & 21.79 (1.24\%) \\
			\hline
			$$         & T=1 & 21.07 & 21.06 (0.02\%) & 21.05 (0.07\%) & 20.93 (0.66\%) & 20.98 (0.41\%) \\
			\hline
			$^{ 30}$P  & T=0 & 12.66 & 12.60 (0.44\%) & 12.55 (0.86\%) & 11.96 (5.86\%) & 11.94 (5.95\%) \\
			\hline
			$$         & T=1 & 11.72 & 11.66 (0.44\%) & 11.62 (0.82\%) & 10.94 (7.11\%) & 10.96 (6.94\%) \\
			
			\hline
			\hline
			
			$^{ 46}$V  & T=1 &  7.92 &  7.92 (0.04\%) &  7.91 (0.10\%) &  7.33 (8.11\%) &  7.76 (2.11\%) \\
			\hline
			$$         & T=0 &  6.93 &  6.93 (0.01\%) &  6.93 (0.07\%) &  6.73 (2.99\%) &  6.79 (2.05\%) \\
			\hline
			$^{ 50}$Mn & T=1 & 12.77 & 12.76 (0.07\%) & 12.75 (0.14\%) & 12.52 (2.02\%) & 12.62 (1.22\%) \\
			\hline
			$$         & T=0 & 12.37 & 12.36 (0.04\%) & 12.34 (0.24\%) & 12.18 (1.61\%) & 12.19 (1.48\%) \\
			\hline
			$^{ 54}$Co & T=1 & 16.14 & 16.12 (0.14\%) & 16.09 (0.28\%) & 15.67 (3.01\%) & 15.86 (1.78\%) \\
			\hline
			$$         & T=0 & 15.93 & 15.92 (0.04\%) & 15.89 (0.22\%) & 15.53 (2.56\%) & 15.66 (1.73\%) \\
			
			\hline
			\hline
			
			$^{106}$I  & T=1 &  5.15 &  5.14 (0.08\%) &  5.13 (0.23\%) &  4.71 (9.37\%) &  4.93 (4.51\%) \\
			\hline
			$$         & T=0 &  4.53 &  4.52 (0.04\%) &  4.51 (0.42\%) &  4.19 (7.84\%) &  4.29 (5.53\%) \\
			\hline
			$^{110}$Cs & T=1 &  8.03 &  7.98 (0.56\%) &  7.97 (0.75\%) &  7.16 (12.14\%) &  7.59 (5.86\%) \\
			\hline
			$$         & T=0 &  7.09 &  7.06 (0.45\%) &  7.04 (0.80\%) &  6.47 (9.64\%) &  6.65 (6.77\%) \\
			\hline
			$^{114}$La & T=1 &  9.76 &  9.72 (0.36\%) &  9.69 (0.73\%) &  8.79 (11.03\%) &  9.27 (5.23\%) \\
			\hline
			$$         & T=0 &  8.95 &  8.93 (0.28\%) &  8.92 (0.42\%) &  8.31 (7.74\%) &  8.51 (5.18\%) \\
			
			\hline
			\hline
			
		\end{tabular}
	\end{center}
\end{table}

With the parameters of the Hamiltonian fixed as explained above, we have studied how
accurate are the energies of the lowest T=0 and T=1 states predicted by the extended QCM
approach for the odd-odd nuclei. The results are presented in Table I. Are  shown the correlation energies 
defined as $E_{corr}=E_0-E$, where $E$ is the total energy while $E_0$ is the non-interacting 
energy obtained by switching off the pairing interactions.
The correlation energies predicted by the QCM functions (6,7) are given in the 4th column. In the brackets
are indicated the errors relative to the exact energies shown in the 3rd column. 
It can be observed that for all the states and nuclei shown in Table I the errors are small, under $1\%$. 
We can thus conclude that the QCM functions (6,7) provide an accurate description of  the 
lowest T=0 and T=1 states of the Hamiltonian (1).

One of the advantages of the QCM approach is the opportunity to study the relevance
of various types of pairing correlations directly through the structure of the trial states (6,7).
As discussed in the previous Section,  this is possible by using  the approximations (10-15).
The correlation energies corresponding 
to these approximations are shown  in Table I. In brackets are given the errors relative to the 
exact results.
One can observe that the smallest errors correspond to  the approximations (10,11) in which the
contribution of the isoscalar pairs in the even-even core of the QCM functions is neglected.
It can be seen that, compared to the calculations with the full QCM  functions,  in these approximations
the errors are increasing by 2-3 times for T=1 states and by larger factors for some T=0 states. 
However, all the errors relative to the exact results remain under $1\%$.

In column 6 are shown the results corresponding to the approximations (12,13) in which
the isovector quartet is taken out from the even-even core. We can see that in this case
the errors are much bigger than in the case when the isoscalar pairs are neglected.
In the last column are given the results of approximations (14,15) obtained by 
neglecting in the QCM states the contribution of the like-particle pairs. It can be noticed
that for all nuclei the states T=1 are better described by a condensate of isosvector pn pairs 
rather than by the approximation (13). On the other hand, the ground T=0 states of $sd$-shell nuclei 
are slightly better described by a condensate of isoscalar pn pairs rather than the approximation  (14). 
However, the latter approximation is by far better than the former in the case
of excited T=0 states of $pf$-shell nuclei and nuclei with $A > 100$.

Overall, these calculations  show that the T=0 and T=1 states cannot
be well described as pure condensates of isoscalar and, respectively, isovector
pairs. In general, by neglecting the contribution of like-particle pairs are generated
large errors. The best  approximation, for both T=0 and T=1 states, is the one in 
which the odd pn pair is appended to a condensate of isovector quartets.
This fact indicates that the  4-body quartet correlations play an important role in odd-odd
N=Z nuclei. As demonstrated in \cite{qcm_iv}, these correlations  are 
missed when the condensate of isovector quartets is replaced by 
products of pair condensates.

\begin{figure}[h]
\centering
\includegraphics[angle=-90,trim = 0.04cm 0.50cm 0.09cm 0.4cm,clip=true, width=15cm]{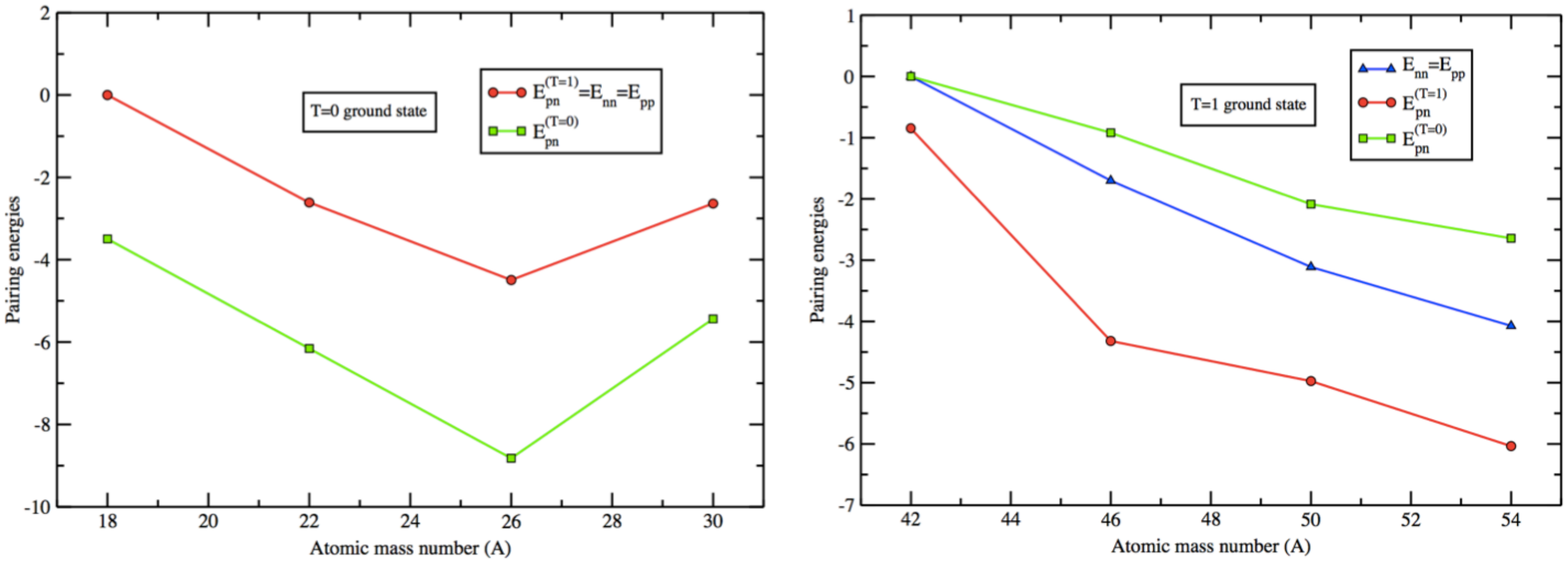}
\caption{Pairing energies, in MeV,  for the odd-odd N=Z nuclei as a function of the mass number A.
In the left (right) panel are shown the results for the $sd$-shell ($pf$-shell) nuclei.}
\end{figure}

For understanding better how the different pairing modes are contributing to the total energy, in 
Fig.2 are shown the isovector and isoscalar pairing energies for the ground states of $sd$ and $pf$
nuclei. The pairing energies are calculated by averaging the corresponding pairing forces on the
QCM functions (6,7). It is important to be observed that the pairing energies for T=1 (T=0) states include also
contributions from the isoscalar (isovector) pairing correlations, a fact which is coming from
the mixing of isovector and isoscalar degrees of freedom  through the even-even core  of the QCM functions.

In the left panel of Fig. 2 are plotted the pairing energies in the ground T=0 states of  $sd$-shell nuclei.
As a reference is shown the pairing energy $E_{pn}^{T=0}$ for $^{18}$F, which corresponds to one
T=0 pair above $^{16}$O. It can be seen that  the curves for $E_{pn}^{T=0}$ and 
$E_{pn}^{T=1}$ are almost parallel. This indicates that the extra pairing energy in the T=0 channel
for $A>18$  is related mainly to the contribution of the odd pn T=0  pairs.  It is also worth noticing that the total pairing energy
in the T=1 channel contains also the contribution from the proton-proton (pp) and neutron-neutron (nn) paring
energies, which, due to the isospin symmetry, are equal to the pn T=1 pairing energy. Therefore, the total
isovector pairing energy is comparable to the isoscalar pairing energy, although the latter contains in addition
a large contribution from the extra odd T=0 pair.

In the right panel of Fig. 2 are plotted  the pairing energies for the  T=1 ground states of $pf$-shell
nuclei. It can be seen that $E^{T=0}_{pn}$ is smaller than $E^{T=1}_{pn}$ and also smaller than
the like-particle pairing energy. At variance with what seen in the left panel, the energy difference  
 $E^{T=1}_{pn}$ -$E^{T=0}_{pn}$ for $A>42$ is much larger than the energy of the odd pn T=1 pair 
 in $^{42}$Sc. Therefore, the larger pn pairing energy in the isovector channel cannot be
 related only  to the extra  pn T=1 pair attached to the even-even core.
  This fact can be traced back to the strong increase of 
 $E^{T=1}_{pn}$ from A=42 to A=46. This increase is mainly related to the contribution, in the 
 nucleus A=46, of the two pn T=1 pairs from the isovector quartet. Since in the isovector quartet all  T=1 pairs 
 have the same structure, the pairing energy of these pn T=1 pairs are equal to the pairing energies of 
 like-particle pairs, which, as seen in A=46, are large, even larger than the energy of the odd pair.

\begin{table}[h]
\caption{ Schmidt numbers for the proton-neutron pairs in the lowest T=1 and T=0 states of
various odd-odd N=Z nuclei. By $K_x$ and  $K_y$ are denoted the Schmidt numbers for the 
pairs $\Gamma^+_0$ and $\Delta^+_0$ while $K_z$  is the Schmidt number for the odd pair, 
i.e., $\tilde{\Gamma}^+_0$ for T=1 states and $\tilde{\Delta}^+_0$ for T=0 states.}
\begin{center}
\begin{tabular}{|c|c|c|c|c|c|c|c|c|c|c|c|c|}
\hline
    
           & \multicolumn{2}{c|} {$^{26}$Al} & \multicolumn{2}{c|} {$^{30}$P} 
           & \multicolumn{2}{c|} {$^{50}$Mn} & \multicolumn{2}{c|} {$^{54}$Co} 
           & \multicolumn{2}{c|} {$^{110}$Cs} & \multicolumn{2}{c|} { $^{114}$La} \\
\hline    
               & T=1      &     T=0        &   T=1  &  T=0      &     T=1    &   T=0         &  T=1   &   T=0        &   T=1   &    T=0           & T=1  &   T=0   \\   
\hline     
$K_x$    &   1.25    &    1.92        &  3.05  & 3.05       &    1.47    &    1.41       & 2.37   &   2.36       &   1.64   &   1.66           & 3.18  &  3.09   \\
\hline 
$K_y$   &   1.97     &    1.31        &  1.89  &   1.56     &     2.39    &    1.33       & 1.72   &  1.25       &   2.24    &   1.88           &  1.16  &  1.24   \\
\hline
$K_z$   &   2.77     &    1.63        &  2.82   &  1.65     &     1.99    &     1.09      &    2.30   &  1.63     &  2.34    &     1.29          &   4.09   &  1.33  \\          
\hline
\end{tabular}
\end{center}
\end{table}

The  T=0 states in odd-odd N=Z nuclei are often  described as states  having a  two quasiparticle
structure. Thus, to evaluate the energies of T=0 states it is commonly employed the blocking procedure,
which means that the odd T=0 pair is not considered as a collective pair in which the nucleons
are scattered on nearby single particle levels but just as a proton and a neutron sitting on a single level.
 In what follows we are going to examine the 
validity of this approximation in the framework of the QCM approach. 
In order to analyze this issue, we need an working definition for the collectivity of a pair.
Here we shall use the so-called Schmidt number, which is commonly employed to analyze
the entanglement of composite systems formed by two parts \cite{Law}. In the case of  a pair operator 
$\Gamma^+=\sum_{i=1}^{n_s} w_i a^+_i a^+_{\bar{i}}$ the Schmidt number  has the
expression  $K=(\sum_i {\omega_i}^2)^2 / \sum_i {\omega_i}^4 $  (for an application
of K to like-particle pairing see Ref. \cite{sandulescu_bertsch}). When there is no
entanglement K=1 while when the entanglement is maximum, which means  equal 
occupancy of all available states, $K=n_s$, where $n_s$ is the number of  states.
As examples, in Table II we show for some nuclei the  Schmidt numbers corresponding 
to the pairs which compose the QCM states (6,7). In Table II by $K_x$ and $K_y$  are denoted the 
Schmidt numbers associated  to the isovector pair $\Gamma^+_0$ and, respectively, to 
the isoscalar pair $\Delta^+_0$. Since in the isovector quartet $A^+$ all the isovector pairs
have the same structure, the like-particle pairs have the Schmidt number $K_x$, as 
the isovector pn pair. By $K_z$ is denoted the  Schmidt number for the odd pair, i.e., 
$\tilde{\Gamma}_0^+$ for T=1 state and $\tilde{\Delta}^+_0$  for the T=0 state. 
We recall that the T=0 state is the ground state for $^{30}$P and excited 
state for $^{54}$Co and $^{114}$La.

From Table II it can be observed that the T=0 pairs are less collective than the
isovector T=1 pairs, which is in agreement with the stronger T=1 pairing correlations
emerging from the results shown in Table I. In particular, the odd T=0 pair
is less collective than the odd T=1 pair. However, in all nuclei, except $^{50}$Mn,
the collectivity of odd T=0 pair is significant and comparable to the collectivity of T=0 pairs in the
even-even core of the QCM states. Therefore these calculations indicate that, in general,
the T=0 states have not a pure two quasiparticle character.

\section{Summary}

In this paper we have  studied the role of isovector and isoscalar pairing correlations in the lowest T=1 and T=0
states of odd-odd N=Z nuclei. This study is performed in the framework of the  QCM approach,
which was extended from the even-even to odd-odd nuclei. In the extended QCM formalism
the  lowest T=0 and T=1 states of odd-odd self-conjugate nuclei are described by a condensate of 
quartets to which is appended an isoscalar or an isovector proton-neutron pair. As in Ref. \cite{qcm_def}, 
the quartets are taken as a linear superposition of an isovector quartet and two collective isoscalar pairs. 
This model was tested for realistic pairing Hamitonians and for nuclei with valence nucleons moving above the cores
$^{16}$O, $^{40}$Ca and $^{100}$Sn. The comparison with exact results shows that the energies 
of the lowest T=1 and T=0 states can be described with high precision  by the QCM approach.  
Taking advantage of the structure of the QCM functions, we have then 
analyzed the competition between the isovector and isoscalar pairing correlations and the  accuracy of various
approximations. This analyze indicates that in the nuclei mentioned above the isoscalar pairing correlations
are weaker but they coexist together with the isovector correlations in both T=0 and T=1 states. 
To describe accurately these states is essential to include the isovector pairing through the isovector
quartets, in which the isovector pn pairs are coupled together to like-particle pairs.
 Any approximations in which the contribution of the like-particle pairing is neglected, including the ones
 in which the T=1 and T=0 states are described by a condensate of isovector pn pairs and, respectively, 
 by a condensate of  isoscalar pn pairs, do not describe accurately the pairing correlations in odd-odd
 N=Z nuclei.
 
In the present study the lowest T=0 and T=1 states are calculated in the intrinsic system of the axially 
deformed mean field and therefore they have not a well-defined angular momentum. The restoration
of  angular momentum  will be treated in a future study.

\newpage
\noindent
{\bf Acknowledgements}
\vskip 0.2cm
\noindent
N.S. thanks the hospitality of IPN-Orsay, Universite Paris-Sud, where this paper was written. 
This work was supported by the Romanian National Authority for Scientific Research
through the grants  PN-III-P4-ID-PCE-2016-048, PN 16420101/2016 and  5/5.2/FAIR-RO.

\end{document}